\documentclass[preprint,12 pt]{article}
\usepackage{arxiv}
\usepackage{graphicx} 
\usepackage{hyperref}
\hypersetup{
	colorlinks=true,
	linkcolor=blue,
	citecolor=blue,
	urlcolor=blue,
}
\usepackage[square,numbers]{natbib}

\usepackage{bm}
\usepackage{amssymb}  

\usepackage{amsthm}
\usepackage{color,soul}
\usepackage{flushend}
\usepackage{mathtools}
\usepackage{xcolor}
\usepackage{array,multirow}
\usepackage[ruled,linesnumbered]{algorithm2e}
\usepackage{siunitx}
\usepackage{subfig}
\usepackage{siunitx}
\usepackage[utf8]{inputenc} 
\usepackage[T1]{fontenc}    
\usepackage{hyperref}       
\usepackage{url}            
\usepackage{booktabs}       
\usepackage{amsfonts}       
\usepackage{nicefrac}       
\usepackage{microtype}      
\usepackage{lipsum}
\usepackage{graphicx}

\begin{document}

%


\title{A Physics-Informed Reinforcement Learning Approach for Degradation-Aware Long-Term Charging Optimization in Batteries}

%


\author{Shanthan Kumar Padisala \thanks{During his part of this work, S. K. Padisala was with The Pennsylvania State University. This does not reflect the view of General Motors.}\\
General Motors Research \& Development\\ Warren, MI 48093, USA.\\ 
\And
Bharatkumar Hegde\\
General Motors Research \& Development\\ Warren, MI 48093, USA.\\
\And
Ibrahim Haskara\\
General Motors Research \& Development\\ Warren, MI 48093, USA.\\
\And
Satadru Dey\\
Department of Mechanical Engineering\\ The Pennsylvania State University\\ University Park, Pennsylvania 16802, USA\\E-mail: skd5685@psu.edu.\\}

\maketitle

\begin{abstract}
Batteries degrade with usage and continuous cycling. This aging is typically reflected through the resistance growth and the capacity fade of battery cells. Over the years, various charging methods have been presented in the literature that proposed current profiles in order to enable optimal, fast, and/or health-conscious charging. However, very few works have attempted to make the ubiquitous Constant Current Constant Voltage (CCCV) charging protocol adaptive to the changing battery health as it cycles. This work aims to address this gap and proposes a framework that optimizes the constant current part of the CCCV protocol adapting to long-term battery degradation. Specifically, a physics-informed Reinforcement Learning (RL) approach has been used that not only estimates a key battery degradation mechanism, namely, Loss of Active Material (LAM), but also adjusts the current magnitude of CCCV as a result of this particular degradation. The proposed framework has been implemented by combining PyBamm, an open-source battery modeling tool, and Stable-baselines where the RL agent was trained using a Proximal Policy Optimization (PPO) network. Simulation results show the potential of the proposed framework for enhancing the widely used CCCV protocol by embedding physics information in RL algorithm. A comparative study of this proposed agent has also been discussed with 2 other charging protocols generated by a non-physics-based RL agent and a constant CCCV for all the cycles. 
\end{abstract}


%

\section{Introduction}


Long charging time of electric vehicles (EVs) compared to the refueling time of internal combustion engine vehicles, has long been a bottleneck in consumer acceptance of EVs. Although high C-rate charging of EV batteries can speed up the charging process, this induces internal failure mechanisms which can potentially degrade the batteries. Furthermore, batteries also experience aging through nominal cycling. Regardless of the cause of the degradation, it is essential to adapt to changing battery health to ensure a longer battery life. This work aims to address this issue by proposing a physics-informed Reinforcement Learning (RL) framework for adapting battery charging protocols with respect to battery age. 

Various battery charging methodologies have been presented in the existing literature. For example, some charging methods are presented based on grey system theory \cite{chen2008design}, ant colony optimization \cite{liu2005search}, and Constant Current Constant Voltage (CCCV) charging \cite{li2020optimized} -- which mostly rely on model-free heuristics. On the other hand, there are methods that leverage physical models of batteries to design charging protocols. For example, optimal control-based approach \cite{klein2011optimal,fang2017health,perez2017optimal2,sattarzadeh2023feedback}, electrochemical model-based approach \cite{yin2021optimal}, and model predictive control based approach \cite{wang2022optimization} have been presented. For detailed reviews on existing battery charging methods, the reader should refer to \cite{bose2022study,chen2021charging}.

More recently, another class of charging methods has emerged in which learning techniques have been used to design the charging strategies. For example, in \cite{aktas2020design}, an adaptive battery charging strategy has been designed, considering the battery temperature. In \cite{korkas2017adaptive}, an adaptive learning strategy was designed that enables optimal dynamic charging of electric vehicle fleets. In \cite{park2022deep}, an reinforcement learning (RL) based framework was developed for the fast charging of batteries. In \cite{yang2022adaptive}, an adaptive control strategy based on RL was developed for a dynamically reconfigurable battery system. Numerous other reports of RL have been published in the literature on the safe charging of batteries \cite{tavakol2024reinforcement,abbasi2024deep}. 

However, most of the aforementioned charging methods attempt to design the charging current profile within a given cycle. Very few works have attempted to make the ubiquitous Constant Current Constant Voltage (CCCV) charging protocol adaptive to changing battery health as it cycles. This work aims to address this gap and proposes a framework that optimizes the constant current part of the CCCV protocol, adapting to long-term battery degradation. Specifically, we formulate a physics-informed Reinforcement Learning (RL) framework that simultaneously performs two tasks: (i) it estimates a key battery degradation mechanism, namely, Loss of Active Material (LAM), and (ii) adjusts the current magnitude of CCCV in accordance with this estimate. An open source battery modeling platform PyBamm \cite{sulzer2021python} in conjunction with the OpenAI Gym-based environment \cite{brockman2016openai} was used to create the battery truth model, while Stable-Baselines-3 \cite{raffin2021stable} has been used to build the RL agent using this truth model as the environment. The rest of the paper details the model formulation, framework development, and simulation results, followed by conclusions.



\section{Dynamics of Battery Systems: Nominal Behavior and Degradation Mechanism}



In this section, we describe the electrochemical models that serve as the truth model for the proposed framework. First, we utilize the Doyle-Fuller-Newman (DFN) model to capture the nominal dynamics of a battery cell \cite{doyle1993modeling}. The system of equations that describe the battery cell in a DFN model is as follows. First, the ionic concentrations inside the spherical active material $C_s^{\pm}$ are described by Fick's law of diffusion:
\begin{align}
    \frac{\partial C_s^{\pm}(r,t)}{\partial t} = \frac{D_s^{\pm}}{r^2} \frac{\partial}{\partial r} \Big( r^2 \frac{\partial C_s^{\pm}(r,t)}{\partial r} \Big), r\in [0,R_s], \label{eq1}
\end{align}
along with the boundary conditions:
\begin{align}
    \frac{\partial C_s^{\pm}}{\partial r}\bigg|_{r=0}=0, -D_s^{\pm}\frac{\partial C_s^{\pm}}{\partial r}\bigg|_{r=R_s} = \frac{j_{Li}^{\pm}}{\alpha_s^{\pm}F}
\end{align}
where the superscript $+$ indicates positive electrode and $-$ indicates negative electrode, $\alpha_s^{\pm}$ are given by $\frac{3\epsilon_s^{\pm}}{R_s^{\pm}}$, with $\epsilon_s^{\pm}$ being the active material volume fractions of the respective electrodes and $D_s^{\pm}$ being the diffusion coefficients of the respective electrode.

The electrolyte phase concentration is given by the following relation:
\begin{align}
    \frac{\partial \big(\varepsilon(x) C_e(x,t)\big)}{\partial t} = \frac{\partial}{\partial x}\Big( D_e^{eff}(x)\frac{\partial C_e}{\partial x}\Big) + \frac{1-t^0_+}{F}j_{Li}(x,t),
    \label{eqn:DFN_Electrolyte}
\end{align}
where $D^{eff}_e$ is the effective diffusion coefficient in the electrolyte, $\varepsilon(x)$ is the porosity in the electrolyte, $t^0_+$ is the transference number of $Li$ ions. Furthermore, $j_{Li}^{\pm}$ is given by the Butler-Volmer kinetics:
\begin{align}
    j_{Li}^{\pm} = \alpha_s^{\pm}i_0\Big(\exp{\frac{\alpha_a F}{RT}\eta} - \exp{\frac{\alpha_c F}{RT}\eta} \Big)
\end{align}
where $\eta^\pm$ is given by $\phi_s^\pm - \phi_e^\pm -U^\pm$, $\phi_s^\pm$ is the corresponding electrode's surface potential, $\phi_e^\pm$ is the electrolyte potential just near the surface of the corresponding electrode, and $U^\pm$ is the open circuit potential. The exchange current $i_0$ is given by:
\begin{align}
    i_0 = k\sqrt{C_e} \sqrt{C^{\pm}_{s_{surf}}}\sqrt{C^{\pm}_{s_{max}} - C^{\pm}_{s_{surf}}} .
\end{align}
Next, the solid phase potential is given by Ohm's law:
\begin{align}
     j^{\pm}_{Li} = \frac{\partial}{\partial x}\Big(\epsilon_s^{\pm}\sigma^{\pm} \frac{\partial \phi_s^{\pm}}{\partial x}\Big),
\end{align}
and the liquid phase potential is given by the following relation:
\begin{align}
    & j_{Li}(x,t) = \nonumber\\
    & \frac{\partial}{\partial x} \Big(k^{eff}(x)\frac{\partial \phi_e}{\partial x} + 2k^{eff}(x)\frac{RT}{F}(1-t^0_+)\frac{\partial \ln C_e}{\partial x} \Big).
    \label{eqn:DFN_Electrolyte_Potential}
\end{align}
The corresponding boundary conditions are given by:
\begin{align}
    -\epsilon_s^{\pm}\sigma^{\pm}\frac{\partial\phi_s^{\pm}}{\partial x}\bigg|_{x=L (or) 0} = \frac{i_{app}}{A_{surf}}, \frac{\partial \phi_s^{\pm}}{\partial x}\bigg|_{x=L-\delta^+  (or) \delta^-} = 0
\end{align}
where, $\sigma^{\pm}$ is conductivity of solid active material, $A_{surf}$ is the electrode plate area, and $i_{app}$ is the applied current. The electrolyte phase potential $\phi_e$ is given by:
\begin{align}
    \frac{\partial}{\partial x} \Big(k^{eff}\frac{\partial \phi_e}{\partial x} + k_D^{eff} \frac{\partial \ln{C_e}}{\partial x}\Big) = -j_{Li},
\end{align}
along with the boundary conditions:
\begin{align}
    \frac{\partial\phi_e}{\partial x} \bigg|_{x=0(or)L} = 0.
\end{align}
Finally, the cell terminal voltage is given by: 
\begin{align}
    V_{cell} = \phi^+_s\bigg|_{x = L} - \phi^-_s\bigg|_{x=0} - \frac{R_{cc}}{A_{surf}}I_{app}(t), \label{eqN}
\end{align}
where $R_{cc}$ is the contact resistance and $I_{app}$ is the applied current. In summary, \eqref{eq1}-\eqref{eqN} capture the nominal dynamics of the battery cell, in terms of it electrochemical behavior.


Next, we discuss the degradation mechanism considered in this work, namely the Loss of Active Material (LAM) in cathode. LAM is a degradation mechanism where the electrochemically active material in either of the electrodes becomes inaccessible or inactive, reducing the cell’s overall capacity. In \cite{zhuo2023degradation}, a degradation model is presented that captures the effects of loss of active material and cyclable lithium on capacity fade \cite{zhuo2023degradation}. In \cite{gao2017lithium}, the capacity fade percentage per cycle as a result of the LAM and the impact of C-rates on this during cycling tests have been modeled. In this work, we leverage this model \cite{gao2017lithium} to capture the capacity fade due to LAM:
\begin{align}
    DS(Q_{loss},I_c) = a(Q_{loss})I_c^{b(Q_{loss})} + c(Q_{loss}), \label{eqn:Degradation}
\end{align}
where $DS$ is the percentage of the capacity fade, which correlates directly to the LAM.

\section{Reinforcement Learning Based Long-Term Charging Optimizer}

\begin{figure}[h!]
    \centering
    \includegraphics[width=0.4\textwidth]{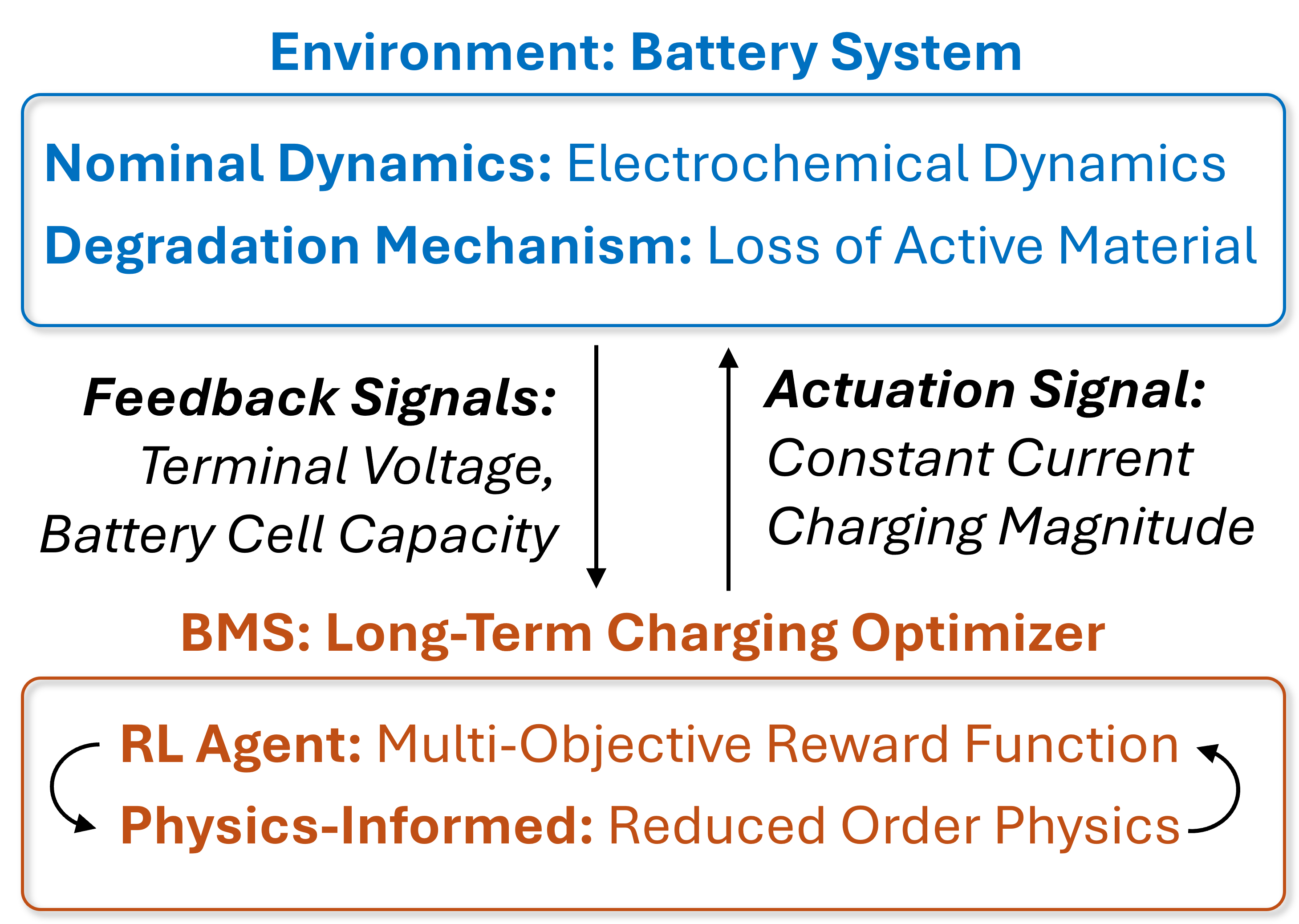}
    \caption{Schematic of the charging optimizer framework.}
    \label{fig:ENV_Architecture}
\end{figure}

In this section, we discuss our proposed long-term charging optimizer framework. The key idea is illustrated in Fig. \ref{fig:ENV_Architecture} where the \textit{environment} (that is, the battery system) and the RL-based optimizer (implemented as an algorithm in the Battery Management System (BMS)) are interacting with each other. The battery system provides information of battery terminal voltage (through direct measurement) and the cell capacity (through periodic monitoring, for example, via reference performance test (RPT)). The optimizer receives this information and deploys an RL agent which in turn performs two simultaneous functions: (i) It estimates the LAM for cathode, and (ii) before the next cycle begins, it adjusts the magnitude of the constant current part of CCCV charging protocol based on this estimate. 

In the training process, the length of each episode is 100 cycles, and during every cycle, the battery undergoes a CC-CV charging with the C-rate dictated by the agent's action. At the beginning of the first cycle, the simulation is initialized with a particular C-rate, and at the end of each cycle, states and rewards are logged. And based upon the states the system propagates through, the agent takes actions that update the C-rates for the next cycle of the 100-cycle-length episode. At the end of the episode, based on the rewards incurred, the agent tunes its parameters so the reward gets maximized, thereby achieving the original objective. Next, we will discuss the design of the RL agent.

We utilize the Markov Decision Process (MDP) formulation \cite{puterman1990markov} for the aforementioned charging problem. MDP-based RL problems are typically defined by system state vector ($s$), action vector ($a$), and reward function ($r$). The goal of the RL agent is to read the states and determine the action vector such that the reward function is maximized. These are given as follows in the context of our problem:
\begin{align}
    \textit{States: } & s = \{ V_{cell}, Q, C_{rate}, \epsilon_s^+\},
\end{align}
where $V_{cell}$ is the cell voltage, $Q$ is the cell capacity, $C_{rate}$ is the constant current magnitude of the CCCV protocol, and $\epsilon_s^+$ is the cathode active material volume fraction capturing the effect of LAM.
\begin{align}    
    \textit{Actions: } & a=\{ \Delta C_{rate}, \Delta\epsilon_s^+\},
\end{align}
where $\Delta C_{rate}$ is the change of $C_{rate}$ and $\Delta\epsilon_s^+$ is the change of $\epsilon_s^+$, respectively, between two consecutive cycles.
\begin{align}    
    \textit{Reward Function: } & r= \alpha_1 r_1 + \alpha_2 r_2 + \alpha_3 ,
\end{align}
where $r_1$, $r_2$, and $r_3$ are three components of the multi-objective reward function with $\alpha_1$, $\alpha_2$, and $\alpha_3$ being their respective weights. Next, we discuss the three components of the reward function. The first reward component $r_1$ is given by:
\begin{align}    
    & r_1 = C_{rate},
\end{align}
which maximizes the current magnitude leading to faster charging. The second reward component $r_2$ is given by:
\begin{align}    
    & r_2 = Q,
\end{align}
which maximizes the cell capacity ultimately reducing capacity loss. Finally, the last reward component $r_3$ is given by:
\begin{align}    
    & r_3 = \frac{1}{N}\sum_{t=0}^N \big[ V_{cell}(t) - V_{r}(t,\epsilon_s^+)\big],
\end{align}
where $t$ is the discrete time instants within a cycle, $N$ is the total cycle time instants, $V_{cell}$ is the measured cell voltage, and $V_{r}$ is the cell voltage predicted by a reduced order physics model. 

In addition to the above reward function, there is also a constraint that updates the reward as follows: if $\epsilon^+_s<0.7\epsilon^+_0$ ($\epsilon^+_0$ is the initial value of the cathode active material volume fraction), then there is a penalty of $500$, thereby updating the reward function to $Reward Function: r = \alpha_1r_1 + \alpha_2r_2 + \alpha_3r_3 -500$. This constraint ensures that the agent does not degrade the battery cells faster.

Referring to the action vector, the action variable $\Delta C_{rate}$ affects the first two reward function components, $r_1$ and $r_2$ -- while the action variable $\Delta\epsilon_s^+$ affects the last reward component $r_3$. Essentially, $\Delta C_{rate}$ dictates the change in current magnitude while $\Delta\epsilon_s^+$ provides an estimate of the parameter $\epsilon_s^+$ capturing LAM degradation.

Essentially, this $r_3$ component of the reward function, as well as the constraint, integrates the physics-information element in the proposed RL framework. This was done by learning the parameter $\Delta\epsilon_s^+$ by utilizing the reduced order physics structure embedded in the model prediction $V_r$. In this case, the predicted voltage $V_{r}$ is generated by using the Single Particle Model (SPM), which captures the reduced order physics of the battery cell \cite{santhanagopalan2006online}:
\begin{align}
    \frac{\partial \theta^{\pm}(r,t)}{\partial t} = \frac{D^{\pm}_\theta} {r^2} \frac{\partial}{\partial r} \Big( r^2 \frac{\partial \theta^{\pm}(r,t)}{\partial r} \Big), r\in [0,R^{\pm}_\theta],
\end{align}
along with the boundary conditions:
\begin{align}
    \frac{\partial \theta^{\pm}}{\partial r}\bigg|_{r=0}=0, -D^{\pm}_\theta \frac{\partial \theta^{\pm}}{\partial r}\bigg|_{r=R_\theta^{\pm}} = \frac{i_{app}}{\alpha^{\pm}_\theta F D^{\pm}_\theta A^{\pm}_\theta L^{\pm}_\theta}
\end{align}
where the superscript $+$ indicates cathode and $-$ indicates anode, $\theta$ is the Lithium concentration, $R_\theta$ is the radius of the particle, $i_{app}$ is the applied current, $\alpha_\theta$ is given by $\frac{3\epsilon^{\pm}_s}{R_\theta}$, where $\epsilon_\theta$ are the active material volume fraction, $D_\theta$ is the diffusion coefficient, $A_\theta$ is current collector surface area, $L_\theta$ is electrode length. Subsequently, the voltage equation for the cell is given by:
\begin{align}
    &V_r = U^+(\theta^+)+\frac{RT}{\alpha^+F}\sinh^{-1}\Big( \frac{i_{app}}{2a_\theta^+ A_\theta L_\theta^+i_0^+} \Big) - \nonumber \\
    &U^-(\theta^-) -\frac{RT}{\alpha^-F}\sinh^{-1}\Big( \frac{i_{app}}{2a_\theta^- A_\theta L_\theta^- i_0^-}\Big) - i_{app}R_f,
\end{align}
where $U^{\pm}$ are the cathode and anode potentials, which are functions of the concentrations $\theta^{\pm}$, and $R_f$ is the contact resistance. 


Next, we detail the RL learning process. After every time step $t$, the RL agent takes action $\Delta C_{{rate}_t}$ which translates the state $\{V_{{cell}_t},Q_t,C_{{rate}_t}\}$ to $\{V_{{cell}_{t+1}},Q_{t+1},C_{{rate}_{t+1}}\}$. The goal of RL training is to find the optimal policy $\pi^{*}_{\lambda}(a|s)$ with policy parameters $\lambda$, which maps the states to actions in order to maximize the expected return \cite{ghadirzadeh2017deep}. The return after the end of an entire episode is given by $G=\sum_{t=0}^{N}\gamma^tr_{t}$ where $G$ is the return value after the entire episode, $r_{t}$ is reward after timestep $t$, and $N$ is the total number of time steps in the episode. 

In this work, Proximal Policy Optimization (PPO) based method has been used for training the RL agent \cite{schulman2017proximal}. PPO is an actor-critic based algorithm, where both these actor and critic networks are learned together. Unlike the traditional policy gradient RL algorithms, instead of directly maximizing the policy gradient objective, PPO introduces a clipped surrogate objective function and maximizes the same. This is computed by a probability ratio given by:
\begin{align}
    & \beta_t =\frac{\pi_\lambda(a_t|s_t)}{\pi_{\lambda_{old}}(a_t|s_t)}, \label{eqn:PropRatio}
\end{align}
where $\pi_\lambda(.)$ and $\pi_{\lambda_{old}}(.)$ are the new and old policies, respectively. If $\beta_t>1$, then the new policy increases the probability of taking the $\Delta C_{rate}$ action $a_t$, while the probability decreases under $\beta_t<1$. PPO modifies the standard objective function into a clipped objective function. The standard objective function is given by the following expected value:
\begin{align}
    & L =E_t\big[ \beta_t \hat{A}_t\big], \label{eqn:StandardObjective}
\end{align}
where $\hat{A}_t = G_t-V_{\phi}(s_t)$ is the advantage function, which is the difference between the return $G_t$ and the value function $V_{\phi}$. This standard objective function is then converted into a clipped version, which is often referred to as the actor network, as described by: 
\begin{align}
    & L^{CLIP}(\lambda) = E\big[\min(R_t(\lambda)\hat{A}_t,\; \text{clip}(R_t(\lambda), 1-\epsilon,1+\epsilon)\hat{A}_t)\big] \label{eqn:Actor}
\end{align}
where $\epsilon$ is a tunable parameter. While the clipped version learns and updates the policy, the critic network learns value function $V_{\phi}(s_t)$, which estimates the expected return starting from the state $s_t$. This critic is given by:
\begin{align}
    & L^{VF}(\phi) = \big( V_{\phi}(s_t)-R_t\big)^2, \label{eqn:Critic}
\end{align}
The resultant objective function that is minimized during the training is:
\begin{align}
    & L(\lambda,\phi) = -L^{CLIP}(\lambda) + c_1L^{VF}(\phi), \label{eqn:PPOLoss}
\end{align}
where $c_1$ is a tunable parameter. Here, the first term $L^{CLIP}(\theta)$ tries to maximize the clipped policy objective (actor) while the second term $L^{VF}(\phi)$ minimizes the value function error (critic). Leveraging the above-mentioned functions, the training process in PPO is given in the Algorithm \ref{table:PPO} \cite{10703056}.


\begin{algorithm}
    \caption{PPO training procedure \cite{10703056}}
    \label{table:PPO}
    \begin{enumerate}
        \item {Initialize policy parameters \textbf{$\theta$}}
        \item {Initialize value function parameters \textbf{$\phi$}}
        \item {\textbf{for} each iteration \textbf{do}}
        \begin{enumerate}
            \item Collect a set of trajectories of the tuples ${s_t,a_t,r_t,s_{t+1}}$ by running the policy $\pi_{\theta}$ in the environment
            \item {\textbf{for} each trajectory \textbf{do}}
            \begin{enumerate}
                \item {\textbf{for} each timestep t \textbf{do}}
                \begin{enumerate}
                    \item Compute advantage estimate $\hat{A}_t$
                \end{enumerate}
                \item {\textbf{end for}}
            \end{enumerate}
            \item {\textbf{end for}}
            \item Minimize the net loss function:\\$L(\lambda,\phi) = -L^{CLIP}(\lambda) + c_1L^{VF}(\phi)$
            \item Update the policy and value function parameters: $\theta$, and $\phi$
        \end{enumerate}
        \item {\textbf{end for}}
    \end{enumerate}
\end{algorithm}

\section{Results and Discussion}
In this section, we discuss some results of the proposed framework. The simulations are performed on a laptop with 16 GB RAM and an Intel i7 $9^{th}$ generation processor (2.6GHz). The environment is designed following the OpenAI Gym API \cite{brockman2016openai}, enabling seamless integration with standard RL algorithms such as PPO \cite{schulman2017proximal}. The primary objective of the environment is to simulate realistic battery behavior under charging while enabling an RL agent to learn optimal policies that account for degradation dynamics. We used the PyBaMM \cite{sulzer2021python} libraries to simulate the DFN model. To simulate LAM degradation mechanism in the cathode, allowing capacity fade to evolve over cycles, we used \ref{eqn:Degradation} with the coefficients $a$, $b$ and $c$ given in Table \ref{tb:Model_Parameters}. 
Each environment step simulates one full charge cycle, from the lower voltage limit of 2.5 V to the upper limit of 4.2 V, under the action-specified C-rate. After each cycle, the state variables are logged. They are given as inputs to the RL agent, which then updates the LAM estimate along with changing the C-rate for the next simulation of the episode.

\begin{table}[ht!]
\centering
\caption{Parameters of the LAM model at different aging conditions \cite{gao2017lithium}}
\label{tb:Model_Parameters}
\begin{tabular}{|c|c|c|c|c|c} 
\hline
     \textbf{ } & \textbf{10\%} & \textbf{20\%} & \textbf{30\%} & \textbf{40\%}\\
     \hline
     \textbf{a} & $0.01058$ & $0.01236$ & $0.01398$ & $0.01566$\\
    \hline
     \textbf{b} & $4.577$ & $3.587$ & $2.938$ & $2.441$\\
    \hline
     \textbf{c} & $0.03375$ & $0.0364$ & $0.03766$ & $0.03806$\\
    \hline
\end{tabular}
\end{table}

In order to benchmark the performance of the proposed framework, we did a comparative study of three charging frameworks: 
\begin{itemize}
    \item \textit{RL with LAM estimate:} This is the proposed framework. Here, the RL agent utilizes the knowledge of physics in terms of the estimation of LAM and decides the constant current magnitude of the CCCV protocol, as discussed in the previous section. Based on the initialization of current magnitude 1.5C, the RL agent outputs change in the C-rate as well as the Cathode Active Material fraction estimate that captures the LAM in cathode.
    \item \textit{CCCV:} Here, the baseline CCCV is used where there is no update of the C-rate after the completion of each cycle and the same C-rate is maintained throughout all the cycles. The C-rate of this algorithm is 1.5C.
    \item \textit{RL without LAM estimate:} Here, the RL agent does not utilize the knowledge of physics to decide the constant current magnitude of the CCCV protocol. Based on the initialization of current magnitude 1.5C, the RL agent outputs only the change in the C-rate. The states, actions, and reward function of this second RL agent are as follows: States: $s = \{V_{cell},C_{rate}\}$, Actions: $a = \{\Delta C_{rate}\}$, Reward Functions: $r = \alpha_1r_1+\alpha_3r_3$
    Due to the absence of the cathode active material volume fraction $\epsilon^+_s$ term, this algorithm does not capture the physics, unlike the first RL algorithm. In order to increase the reward, the algorithm keeps on increasing the C-rate due to the missing cell capacity and degradation physics.
\end{itemize}
These frameworks have been tested with a 100-cycle battery lifespan. Next, we discuss the training process of the RL frameworks mentioned above. The duration of an entire episode is 100 cycles. Each step of this episode corresponds to a full cycle of the CCCV charging scenario. During every step, CCCV charging is performed until the CV current drops to 0.1 A, after which the charging current is cut off. After the end of each cycle, the state transitions result in a reward depending on how close the target is to the goal. The agent then takes actions according to the new states, using which the next step of the episode is simulated. This continues till the end of the episode. If the system goes out of the defined bounds, the episode ends early with a large penalty. After every few iterations, the network's weights are updated to take actions that result in increasing reward.



\begin{figure}[h!]
    \centering
    \includegraphics[width=0.5\textwidth]{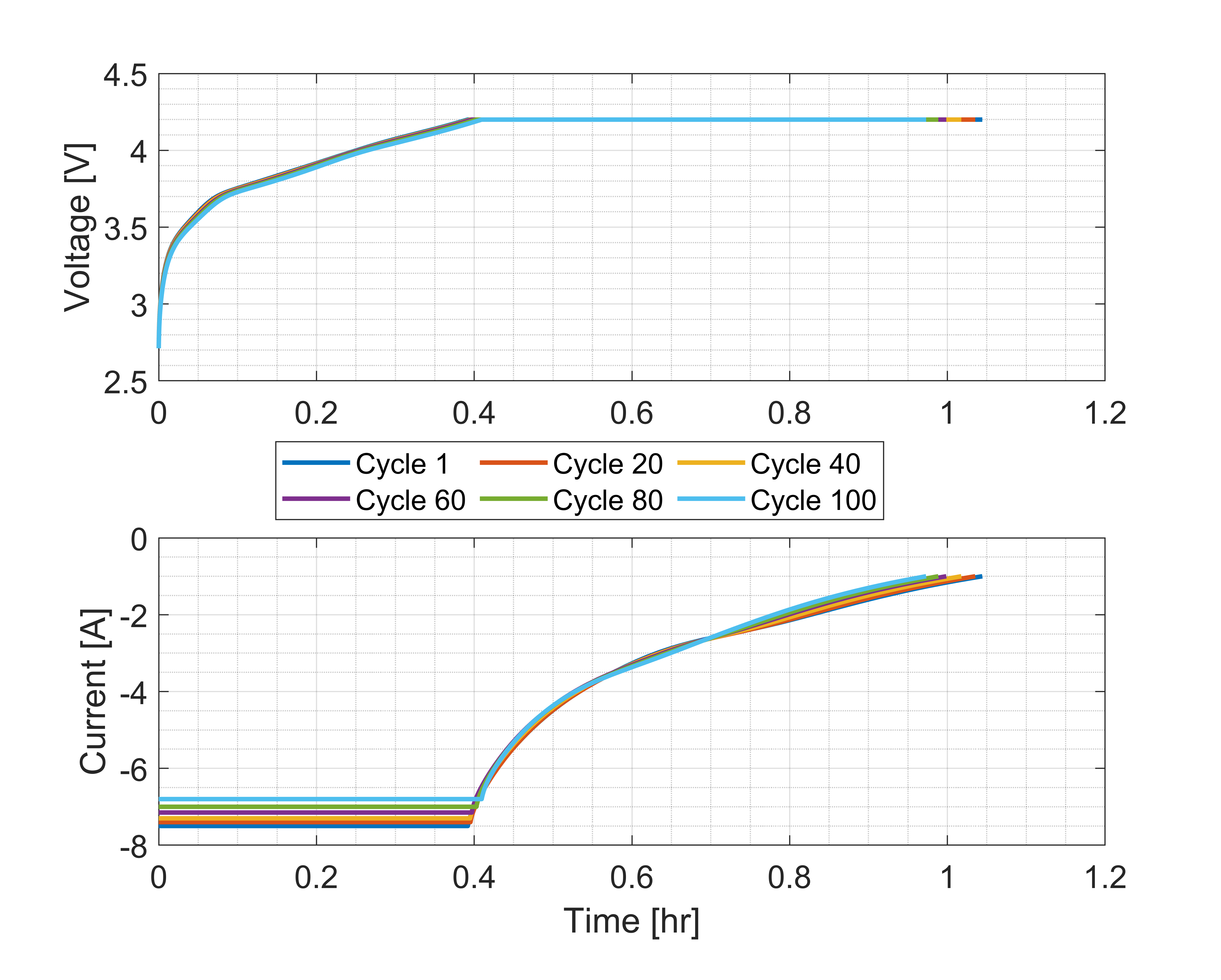}
    \caption{Applied charging current and corresponding terminal voltage responses under various points of battery age.}
    \label{fig:Results01}
\end{figure}

\begin{figure}[h!]
    \centering
    \includegraphics[width=0.5\textwidth]{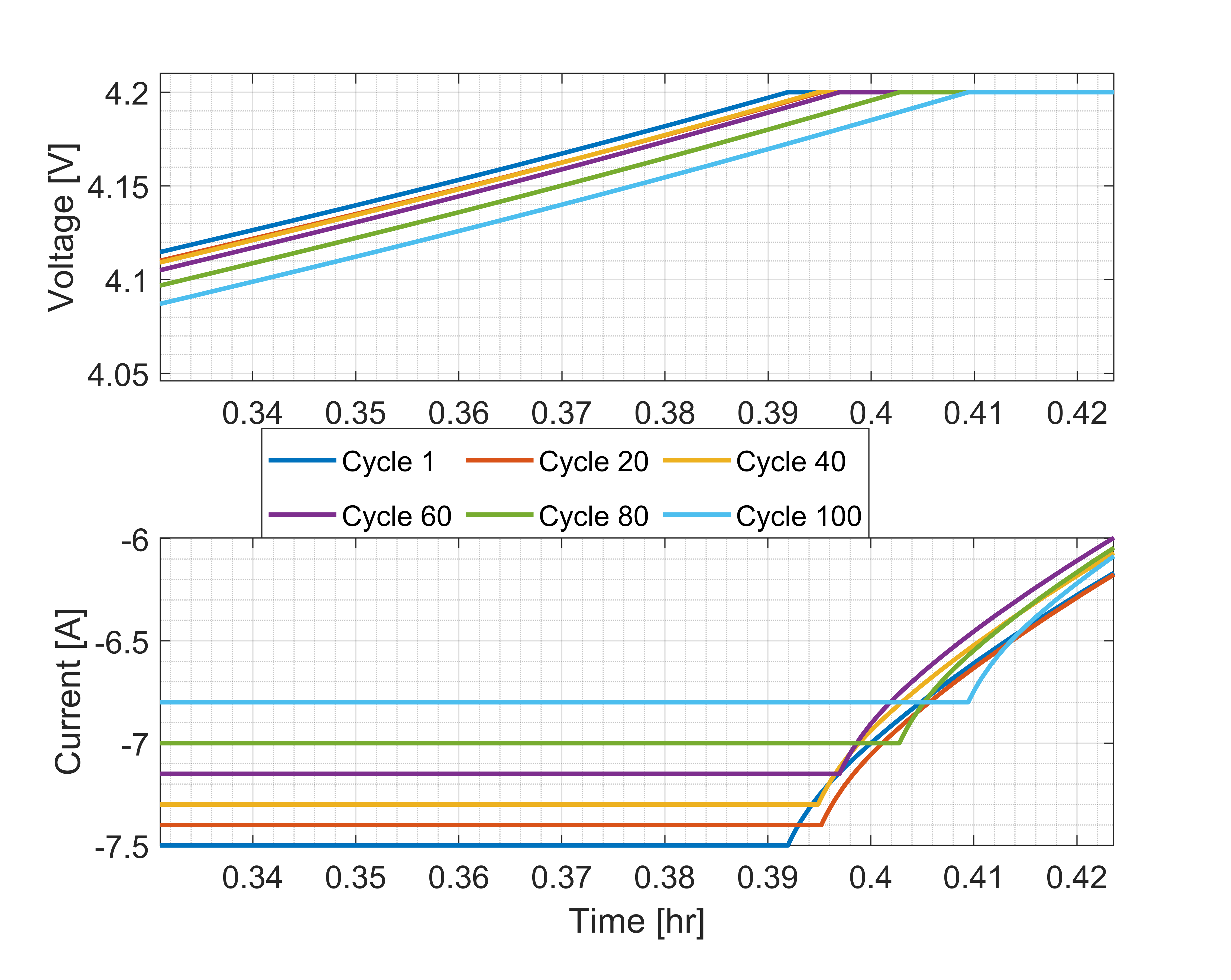}
    \caption{Changes in applied charging current and corresponding changes in terminal voltage responses under various points of battery age.}
    \label{fig:Results02}
\end{figure}

\begin{figure}[h!]
    \centering
    \includegraphics[width=0.5\textwidth]{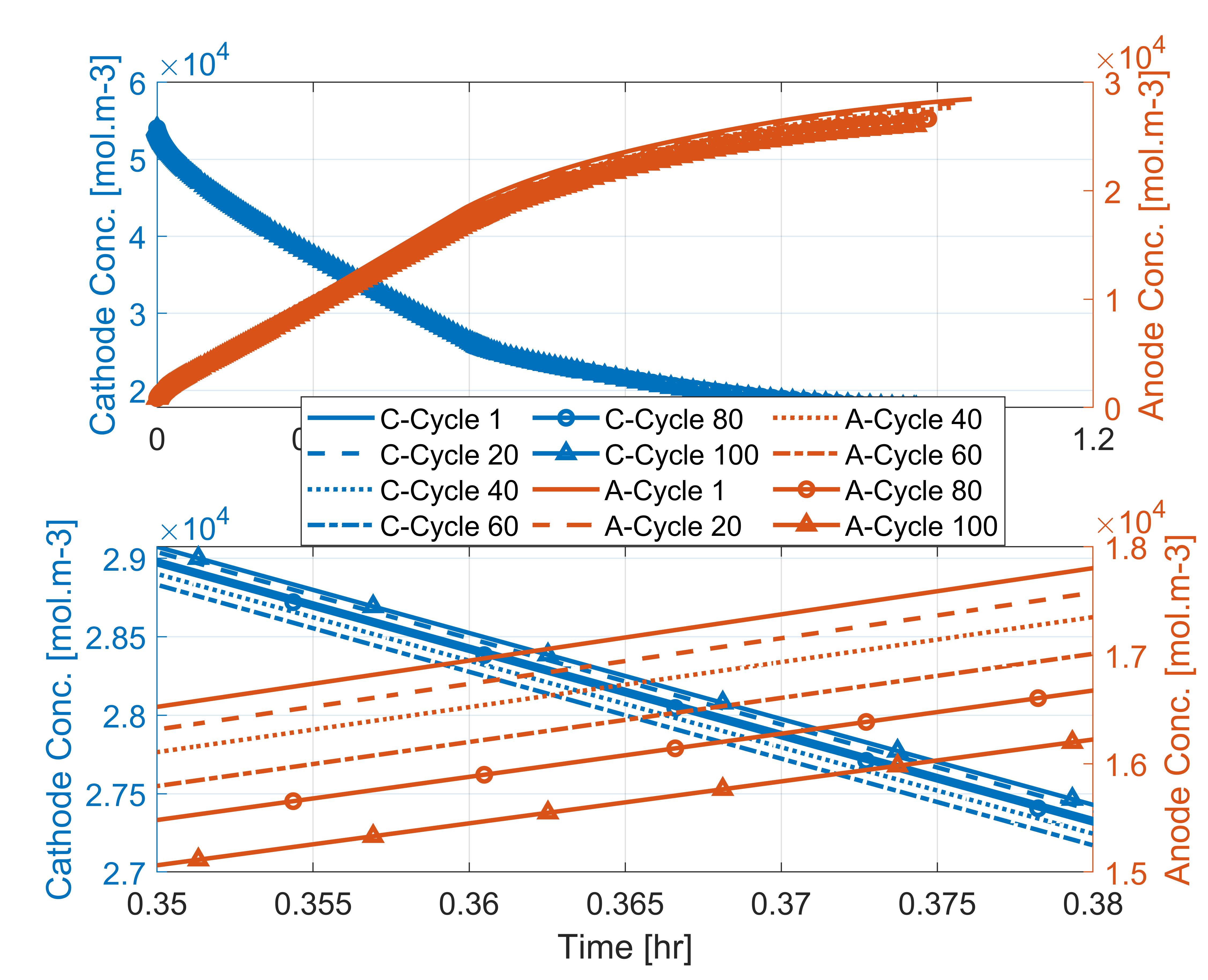}
    \caption{\textit{Top Plot:} Lithium concentrations in anode and cathode during charging under various points of battery age. \textit{Bottom Plot:} Zoomed-in Lithium concentrations in anode and cathode during charging under various points of battery age.}
    \label{fig:Results03}
\end{figure}

\begin{figure}[h!]
    \centering
    \includegraphics[scale = 0.7]{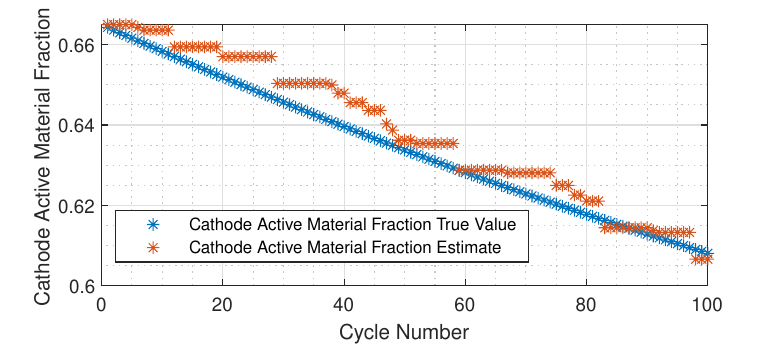}
    \caption{The estimation performance by the proposed framework (\textit{RL with LAM estimate}): Comparison of the true cathode active material volume fraction and the estimates by the RL agent.}
    \label{fig:Results_Algo1}
\end{figure}

First, we illustrate the results of \textit{RL with LAM estimate}. In Fig. \ref{fig:Results01}, the applied charging current and the corresponding terminal voltage responses are shown for cycle \# 1, 20, 40, 60, 80, and 100 -- while a zoomed-in version of the same responses is illustrated in Fig. \ref{fig:Results02}. As we can see, the constant current part of the CCCV charging current has been reduced as the battery experienced more cycles, responding to the LAM degradation of the battery cathode. Consequently, the charging time has also increased due to the reduced magnitude of the charging current, captured by the fact that the terminal voltages hit the 4.2 V limit with a slower time window as the battery ages. Next, in Fig. \ref{fig:Results03}, we show the internal battery states in terms of the Lithium concentration responses in cathode and anode during charging. As expected, the anode concentration trajectories are lowered in magnitudes at the end of the charging as the battery ages, while the cathode concentration trajectories converged to higher magnitudes -- accounting for the degradation in the battery. Figure \ref{fig:Results_Algo1} compares the true value of the cathode's active material volume fraction and the corresponding estimates made by the RL agent in the proposed framework. It can be seen that these estimates are reasonably close to the true values, providing an important physics insights to the RL component that dictates the charging current. 

\begin{figure}[h!]
    \centering
    \includegraphics[scale = 0.7]{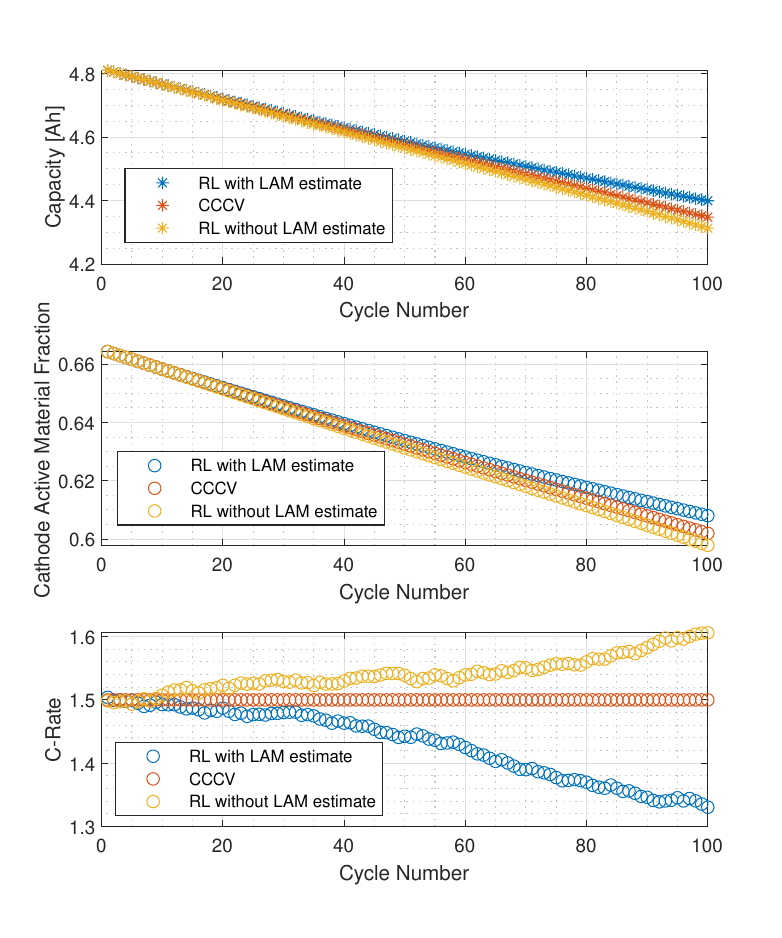}
    \caption{Comparison of the proposed framework with two other benchmarks: CCCV and RL without LAM estimate. \textit{Top Plot:} Comparison of cell capacity degradation under 100-cycle battery lifespan. \textit{Middle Plot:} Comparison of cathode active material volume fraction under 100-cycle battery lifespan. \textit{Bottom Plot:} Comparison of constant current charging C-rates under 100-cycle battery lifespan.}
    \label{fig:Results1}
\end{figure}

Finally, Fig. \ref{fig:Results1} compares the performance of the proposed framework with respect to the two other benchmarks mentioned above. Three key battery quantities are considered for comparison: cell capacity, cathode's active material volume fraction, and charging C-rate for the constant current part. The top plot of Fig. \ref{fig:Results1} illustrates the reduction in cell capacity $Q$, where it can be clearly seen that the proposed framework results in the minimum capacity reduction. This lower degradation in the proposed framework is also reflected in the minimum cathode active material fraction drop as shown in the middle plot of Fig. \ref{fig:Results1}. While Algorithm 3 which has no knowledge of physics and tries to fast charge the battery cells, showcases the maximum degradation. The charging current output (in terms of C-rates) of the three frameworks are shown in Fig. \ref{fig:Results1}. It is clear that the proposed framework learns the physics and adjusts the C-rates to reduce the battery degradation, however, \textit{RL without LAM estimate} attempts to increase the C-rate to fast charge the battery being unaware of the battery degradation. On the other hand, CCCV baseline kept the constant C-rate throughout the lifespan of the battery. It is observed that at the end of 100 cycles, proposed framework results in 8.54\% of capacity fade, CCCV baseline results in 9.61 \%, and \textit{RL without LAM estimate} results in 10.34 \% fade. We can also observe an interesting result that utilizing RL without any physical insights can potentially lead to worse degradation compared to even the CCCV baseline. This underscores the importance of using physical insights in data-driven charging optimization in battery systems.


\section{Conclusion}

In this work, a physics-informed RL approach has been presented to adapt the CCCV charging protocols with battery aging. The physics information was embedded in the RL training by learning the cathode active material volume fraction. The reward function for RL training was designed to capture the multi-objective nature of the problem, including components for fast charging, lowering capacity loss, and learning cathode active material volume fraction. It has been found that compared to a baseline constant CCCV protocol and a standard RL algorithm without physics information, the physics-informed RL algorithm results in reduced degradation as it adopts the charging rates according to the degradation.

\bibliographystyle{IEEEtran}
\bibliography{ref,reference_1}

\end{document}